\documentclass[12pt]{article}  
\newcommand{\comma}{\;\; , \; \; }
\newcommand{\period}{\;\; .}
\newcommand{\eq}{\; = \;}
\newcommand{\sep}{\;\; , \;\;}

\newcommand{\be}{\begin{equation}}
\newcommand{\bd}{\begin{displaymath}}
\newcommand{\ee}{\end{equation}}
\newcommand{\ed}{\end{displaymath}}
\newcommand{\ba}{\begin{eqnarray}}
\newcommand{\ea}{\end{eqnarray}}

\renewcommand{\i}{{\rm i}}

\newcommand{\e}{{\rm e}}

\renewcommand{\arraystretch}{1.5}

\newcommand{\W}{\overline{\cal W}}
\newcommand{\Ws}{{\cal{W}}}
\newcommand{\p}{q}
\newcommand{\q}{p}
\newcommand{\Z}{{\cal Z}}
\renewcommand{\i}{{\rm i}}

\title{A conjecture for  the superintegrable chiral Potts model}

\author{ R.J. Baxter\\
{\protect \small  Mathematical
Sciences Institute,  The Australian National}\\
{\protect  \small  University,
 Canberra, A.C.T. 0200, Australia, \small e-mail: none }}

\date{}

\begin{document}


\maketitle


 \abstract{We adapt our previous results for the ``partition function'' 
of the superintegrable chiral Potts model with open boundaries to 
obtain the  corresponding matrix elements of $\e^{-\alpha H}$, where 
$H$ is the associated hamiltonian. The spontaneous magnetization
${\cal M}_r$ can be expressed in terms of particular  matrix elements of 
$\e^{-\alpha H} S^r_1 \e^{-\beta H}$, where $S_1$ is a diagonal matrix.
We present a conjecture for these matrix elements as an $m$ by $m$ 
determinant, where $m$ is proportional to the width of the lattice. The 
author has  previously derived
the spontaneous magnetization of the chiral Potts model by 
analytic means, but hopes that this work will facilitate 
a more algebraic derivation, similar to that of Yang for the Ising model.}








 \vspace{5mm}




 \section{Introduction}


\setcounter{equation}{0}

In a previous paper\cite{Baxter1989}, we obtained the partition function 
$\tilde{Z}_Q$ (here referred to as $\tilde{Z}_{\q}$) of the superintegrable 
chiral  Potts model with open 
boundary   conditions. It is a simple product of elements of two-by-two 
 matrices, reflecting the fact that there is a reduced representation 
in which the transfer matrices have a direct product  
structure, similar to that of the Ising model.\cite{ Tarasov1991}

Very recently, we have considered the problem of calculating the 
spontaneous magnetization $\cal M$ of the square lattice Ising 
model.\cite{Baxter2008}  We 
used the method of Yang\cite{Yang1952} and defined  $\cal M$ 
in terms of the partition function $\widetilde{W}$ on a cylindrical 
lattice of $L$ columns with fixed-spin boundary conditions 
on the upper and lower rows, with a single-spin operator $S_1$ acting 
on a spin located within the lattice. For convenience, we took the
limit when the transfer matrix could be replaced by the exponential of 
an associated hamiltonian.

The Clifford algebra technique of Kaufman\cite{Kaufman1949} can
 still be applied to 
this system, so that $\widetilde{W}$ can be calculated as the 
square root of an $L$-dimensional determinant. This can be  further
reduced to a determinant (without the square root) of dimension
approximately $L/2$.


Here we write down corresponding definitions of $\widetilde{W}$ for the 
 $N$-state superintegrable chiral  Potts model, which reduces to the 
 Ising model when $N=2$. 
 We conjecture in (\ref{cnjW}) - (\ref{conj2a})  that  $\widetilde{W}$ is
 also given by a determinant of dimension smaller than $L$, being
 a fairly immediate  generalization
 of that for the Ising case. If true, this is an exact formula for finite
 lattices, containing three additional arbitrary parameters
 $\alpha, \beta , x$ in addition to $N, L$ and the labels $\q, \p$ of the
 appropriate sub-spaces. It is therefore easy to test numerically, and 
we have tested it to 60 or more digits of accuracy
 for various small values of $N, L$ (up to $N+L = 10$).
 
If this conjecture is indeed true, then the
 spontaneous magnetization of the superintegrable chiral  Potts model
 is given by the expression (\ref{resultM}) below.
 This necessitates taking the limit $L \rightarrow \infty$ . As yet we 
 have not done this,
 but we have observed numerically that  (\ref{resultM}) does indeed 
 appear to converge to the known result (\ref{Albconj}). The author 
 has previously derived  (\ref{resultM}) by analytic
 methods\cite{Baxter2005a,Baxter2005b} that apply in  the large-lattice 
 limit, but it would still be interesting to have an
 algebraic derivation that could give greater insight into the properties
 of the model on a finite lattice.
 
 \section{Partition function}


\setcounter{equation}{0}
\subsection*{Definition}

 The model is defined on the square lattice, rotated through 
$45^{\circ}$, with $M+1$ horizontal rows, each containing  $L$ spins, 
as in Fig. 1.


 \setlength{\unitlength}{1pt}
 \begin{figure}[hbt]
 \begin{picture}(400,160) (-23,17)

 \put (43,43) {\line(1,1) {74}}
 
 \put (123,63) {\line(1,1) {54}}
 \put (183,63) {\line(1,1) {54}}
 \put (243,63) {\line(1,1) {43}}
 \put (63,3) {\line(1,1) {54}}
 \put (123,3) {\line(1,1) {54}}
 \put (183,3) {\line(1,1) {54}}
 \put (243,3) {\line(1,1) {43}}
 \put (43,104) {\line(1,1) {14}}
 
 \put (63,117) {\line(1,-1) {54}}
 \put (123,117) {\line(1,-1) {54}}
 \put (183,117) {\line(1,-1) {54}}
 \put (243,117) {\line(1,-1) {43}}
   
 \put (43,77) {\line(1,-1) {69}}
 \put (123,57) {\line(1,-1) {54}}
 \put (183,57) {\line(1,-1) {54}}
 \put (243,57) {\line(1,-1) {43}}
  \put (43,16) {\line(1,-1) {14}}

 \multiput(60,0)(60,0){4}{\circle{7}}
 \multiput(60,60)(60,0){4}{\circle*{7}}
 \multiput(60,120)(60,0){4}{\circle{7}}
 \multiput(90,30)(60,0){4}{\circle*{7}}
 \multiput(90,90)(60,0){4}{\circle*{7}}

   \put (51,-13) {$a$}
 \put (111,-13) {$a$}
 \put (170,-13) {$a$}
 \put (230,-13) {$a$}
 
   \put (86,14) {$1$}
   \put (146,14) {$2$}
   \put (266,14) {$L$}

   \put (61,128) {$0$}
    \put (121,128) {$0$}
   \put (181,128) {$0$}
   \put (241,128) {$0$}

 \put (175,45) {$i$}
 \put (205,75) {$j$}
  \put (181,73) {$\Ws$}
  
  \put (314,-4) {$1$}
  \put (314,26) {$2$}
  \put (314,116) {$M+1$}
 \put (107,74) {${\W}$}
 
 \end{picture}
 \vspace{1.5cm}
 \caption{\footnotesize The square lattice $\cal L$  turned 
 through $45^{\circ}$.}
 
  \label{sqlatt45}
 \end{figure}
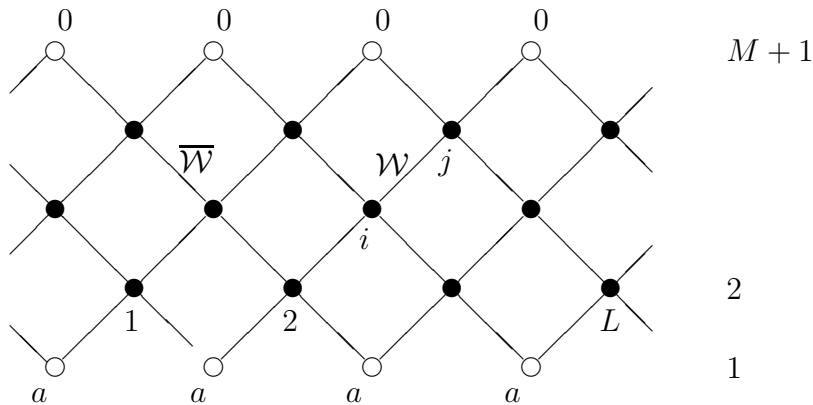


 We impose cylindrical boundary 
 conditions, so that the last column $L$ is followed by the first column 
 1. At each site  $i$ there is a spin $\sigma_i$, taking the values
 $0, 1, \dots, N-1$. The spins in the bottom row are fixed to have 
 value $a$, those in the top row to have value 0. Adjacent spins  
 $\sigma_i, \sigma_j$ on southwest to northeast edges (with $i$ below 
 $j$) interact with Boltzmann weight ${\Ws}(\sigma_i - \sigma_j)$;  
 those on southeast to northwest edges with weight 
 $\W(\sigma_i - \sigma_j)$.

 These $\Ws, \W$ are the Boltzmann weight functions:
 \ba   {\Ws} (n) = & {\Ws}  (n+N) = &  \mu^n  \prod_{j=1}^{n}   
 (1-\omega^j y) /(1- \omega^j x ) 
 \comma \nonumber \\
  {\W}(n)  = & \W(n+N) = &  \mu^{-n}  \prod_{j=1}^{n}  
   (\omega-\omega^j x) /(1- \omega^j y ) \comma \ea
  where  $\omega = \e^{2 \pi \i /n}$,  $x$ and $y$  are complex 
  parameters, and
   \be \mu^N = (x^N-1)/(y^N-1) \period \ee
   
  An important associated parameter is
 \be k' \eq (x^N-1)(y^N-1)/(y^N-x^N) \period \ee
   The partition function, which depends on $a$,  is 
   \be Z_a  \eq \sum_{\sigma} \prod_{\langle i,j \rangle} 
  \Ws(\sigma_i - \sigma_j )  \prod_{\langle i,j \rangle} 
  \W(\sigma_i - \sigma_j ) \comma \ee
   the products being over all edges of the two types. The sum is 
  over all values of all the free spins.
    The partition function  can be written as
   \be Z_a \eq u_a^{\dagger} \, { T} ^M \, u_0 \comma \ee
   where ${ T}$ is the row-to-row transfer matrix, with elements
   \be 
   T_{\sigma, \sigma'} \eq  \prod_{i=1}^L  \Ws(\sigma_i - \sigma'_{i+1} ) 
  \W(\sigma_i - \sigma'_{i} )  \comma  \ee
   $\sigma$ being the set of all spins $\sigma_1, \ldots , \sigma_L$ 
  in one row, and $\sigma'$ being the set in the row above. Thus  
  $ T$ is an $N^L$ by $N^L$ matrix. The vector $u_a$ is of 
  dimension $N^L$, with entries
   \ba \label{defua}
   ( u_a)_{\sigma}  & = & 1\;\;  {\rm if }  \;\; \sigma_1 = 
   \cdots  = \sigma_L = a \comma \nonumber \\
     & = & 0 \; \; {\rm otherwise} \period \ea


 The superintegrable chiral Potts model is a special case of the more
 general solvable chiral Potts model, which satisfies the star-triangle 
 relation.\cite{BPY1988} This ensures that two transfer matrices $T, T'$, 
 with different values of $x, y$, but the same value of $k'$,
 commute.
  
    
    
    \subsection*{The spin-increment matrix $R$}


  Let $R$ be the $N^L$ by $N^L$ matrix with entries
   \be R_{\sigma, \sigma'} \eq  \prod_{j=1}^L  
 \delta (\sigma_j , \sigma'_j + 1) \comma   \ee
   where $\delta(a,b) = 1 \;   {\rm if} \;a = b \; 
({\rm modulo} \;  N)$,  else  $\delta(a,b) = 0$.
   Then pre-multiplying by $R$ has the effect of increasing all 
 spins by 1 (modulo $N$),  hence $R u_a = u_{a\scriptscriptstyle{+1}}$ 
and $R$ commutes with $ T$:
 \be \label{comm} R \, { T} = { T } R \period \ee
 For this reason it is natural to use the Fourier transform of $u_a$:
 \be \label{vu}
 v_{\q} \eq   N^{-1/2}  \; \sum_{a=0}^{N-1} 
\omega^{- a \q} \, u_a \period \ee
 taking $\q = 0, \ldots, N-1$. This $\q$ replaces the $Q$ of
 \cite{Baxter1988,Baxter1989}. Then
 \be \label{Rmult}
 R\, v_{\q}  = \omega^{\q}  v_{\q}   \period \ee

  If we also define
 \be \label{ZtZ}
\tilde{Z}_{\q}  \eq \sum_{a=0}^{N-1} \, \omega^{{\q}  a} \, Z_a \comma 
 \ee
 then
 \ba \tilde{Z}_{\q}    & = &  N^{1/2} \, v_{\q} ^{\dagger} T^M u_0 \nonumber \\
 & = & \sum_{\p =0}^{N-1}  v_{\q} ^{\dagger} T^M v_{\p }  \period \ea
 {From} (\ref{comm}) , we can replace $T^M$ in the summand by 
 $R^{-1}T^M R$, and from (\ref{Rmult}) this is equivalent to multiplying
 by $\omega^{\p - {\q} }$. This in turn means the summand must vanish 
 unless $ \p = \q$, so 
  \be \tilde{Z}_{\q}  =  v_{\q} ^{\dagger} T^M v_{{\q} }  \period \ee




\subsection*{The sub-space $V_p$}


 Following the observations of Albertini {\it et al}\cite{Albertini1989},
 we showed in refs.\cite{Baxter1988,Baxter1989} that if one operates
 on $v_{\q} $ by any product of matrices $T$, with different values of 
 $x,y$ but the same value of $k'$,  then all the  vectors generated 
 lie in a vector space $V_{\q} $, where ${\q} =0, \ldots , N-1$.  
 For any vector $v$ in $V_{\q} $,
 \be  \label{RVq}
  R \, v \eq \omega^{\q}   v \period \ee
 
 We also showed that the transfer matrices satisfied a functional 
 relation that determined their eigenvalues, and derived the 
 result (\ref{Zq}) for the partition function  $\tilde{Z}_{\q}$. 
 
 If
 \be \label{defm} m = m_{\q}  = {\rm integer \; part \; of\; }  
 \left[\frac{(N-1) L - {\q} }{N} \right] \comma  \ee
 then there are just $2^m$ distinct eigenvalues. What we have not 
 shown, but believe to be true, is that each such eigenvalue occurs
  just once, so that $V_{\q} $ is of dimension $2^m$. Certainly, by 
  continuity from the case $k' = 0$,  the largest
  eigenvalue (which is the one we most often consider) occurs just once.
  
  For the case when $\q = 0$ and $L$ divides by $N$, Au-Yang and Perk
  have recently obtained the eigenvectors explicitly.\cite{AuYangPerk2008}
 
Two vectors $v, w$ in different spaces $V_p$, $V_{{\q} }$ (with 
$\p \neq {\q} $) are necessarily orthogonal, i.e.  $v^{\dagger}{\cdot} \, w = 0$.


Define 
\be \label{defP}
 P(z^N) \eq z^{-{\q} } \, \sum_{n=0}^{N-1} 
\omega^{(L+{\q} )n} {(z^N-1)/(z-\omega^n)}^L \period \ee
Then $P(w) = P_{\q} (w) $ is a polynomial in $w$ of degree $m$. Let 
its zeros be
$w_1, \ldots ,w_m$ and define
$\theta_1, \ldots ,\theta_m$  $ =  \theta ({\q} ,1), \ldots ,\theta ({\q} ,m)$  by
\be \label{deftheta}
\cos \theta_j \eq \cos [ \theta ({\q} ,j) ] \eq  (1+w_j)/(1-w_j) \sep 0 
< \theta_i < \pi  \comma \ee
for $j = 1,\ldots, m $. Set 
\be G \eq (x^N y^N-1)/(y^N-x^N) \comma \ee
\be g \eq N(1-x^{-1})/(1-x^{-N}) \comma \ee
define the two-by-two matrices 
\renewcommand{\arraystretch}{1.2}
\setlength{\arraycolsep}{4pt}

\be  S = \left( \begin{array}
{cc} 1 & 0 \\
0  & -1 \end{array} \right) \sep
C = \left( \begin{array}
{cc} 0 & 1 \\ 1 & 0 \end{array} \right) \comma \ee
\be  F(x,y,\theta) \eq \frac{1-x^{-N}}{2 k'} \left[ G \, I_2 +
(1-k' \cos \theta )S -k' \sin \theta \, C \right] \comma \ee
$I_2$ being the identity matrix, and set
\renewcommand{\arraystretch}{1.2}
\setlength{\arraycolsep}{4pt}

\be \label{defD}
D(\cos \theta ) \eq  (
\begin{array} {c c} 1 & 0 \end{array} ) \, {\displaystyle{\cdot}}
\,  F(x,y,\theta)^{\raisebox{0.8ex}{\it \scriptsize M}} 
{\displaystyle{\cdot}} \left( \begin{array} {c} 1 \\ 0 \end{array}
 \right)  \ee
\renewcommand{\arraystretch}{1.2}
then in \cite{Baxter1989} we find that
\be \label{Zq}
\tilde{Z}_{\q}  \eq g^{L M} x^{- M {\q} } \, D(\cos \theta_1) \,  
D(\cos \theta_2) \cdots  D(\cos \theta_m) \period \ee




\section{The hamiltonian limit}


\setcounter{equation}{0}


Take \be \mu  = \e^{- 2 \epsilon } \ee
and consider the limit when $\epsilon \rightarrow 0 $. Then to 
first  order in $\epsilon$
\be x = y =  1+ 2 k' \epsilon \comma \ee
\be {\Ws}(n) = 1 - 2 n \epsilon \sep
\W (n) = 2 k' \epsilon /(1-\omega^{-n})  \ee
for $ 0  < n < N$, while ${\Ws}(0) = \W (0) = 1$. Noting that
\be N-1-2j = 2 \sum_{n=1}^{N-1} \frac{\omega^{n j}}{1-\omega^{-n} }   
 \ee
for $ 0 \leq j < N $, it follows that
\be \label{Tlim}
T \eq [1 - (N-1) L \epsilon] \, I -\epsilon \, {\cal H} \comma  \ee
where $I$ is the identity matrix and 
\be {\cal H} = -2 \sum_{j=1}^L \sum_{n=1}^{N-1} ( {\Z}_j^n {\Z}_{j+1}^{-n} + 
k' X_j^n)/(1-\omega^{-n}) \period \ee
This is the hamiltonian associated with the transfer matrix $T$.
Since all transfer matrices with the same value of $k'$ commute,
they also commute with $\cal H$.
Here ${\Z}_j, X_j$ are the $N^L$ by $N^L$ matrices of \cite{Albertini1989},
 with elements
\be \left( {\Z}_j \right)_{\sigma, \sigma'} = \omega^{\sigma_j} \, 
\prod_{m=1}^L \delta(\sigma_m,\sigma'_m) \comma \ee
\be \left( X_j \right)_{\sigma, \sigma'} =  \delta(\sigma_j,\sigma'_j+1)
{\prod_{n=1}^L}^{\! \raisebox{-10pt}{*}} \delta(\sigma_n,\sigma'_n) 
 \comma \ee
the $*$ on the last product indicating that that it excludes
the case $n=j$.

The hamiltonian $\cal H$ is known to have very special properties. In 
particular Au-Yang and Perk showed that it satisfies the ``Onsager 
algebra''.\cite{AuYangPerk1989}

Still working to first order in $\epsilon$, we obtain
\bd g = 1+(N-1) k' \epsilon \comma \ed
\bd \frac{(1-x^{-N}) G}{2 k'} \eq 1 -   N (1+k')\epsilon \comma \ed
\bd F(x,y,\theta)  = 
 [1- N (1+k') \epsilon] I_2 + N \epsilon
\left[ (1-k' \cos \theta) S - k' \sin\ \theta \, C \right] \period
 \ed


\noindent Now we take 
\be  \epsilon =  \alpha/M \comma \ee
and let $M \rightarrow \infty$, keeping $\alpha$ fixed. Then 
\be F(x,y, \theta)^M \rightarrow 
\exp \{ N \alpha [-(1+k' ) I_2 + (1-k' \cos \theta )S -k' \sin \theta
 \, C ] \} \ee
and from (\ref{Tlim}),
\be T^M \rightarrow  \e^{-(N-1)L \alpha } \, \exp ( - \alpha  \cal H )
 \period \ee

{From} (\ref{defD}) and (\ref{Zq}), it follows that
\be \label{resvq}
v_{\q} ^{\dagger} \exp(-\alpha {\cal H} ) \, v_{\q}  \eq  e^{-\mu \alpha }
\, \overline{D} (\cos \theta_1) \cdots \overline{D} (\cos \theta_m)
\comma \ee
where
\be \label{defmuq}
 \mu = \mu_{\q}  = 2 k' {\q}  +(1+k')(mN-NL+L) \comma \ee
\be \label{defoD}
\overline{D}(\cos \theta ) \eq  (
\begin{array} {c c} 1 & 0 \end{array} ) \, {\displaystyle{\cdot}}
\, \exp[- \alpha \tilde{F}(\theta) ] 
{\displaystyle{\cdot}} \left( \begin{array} {c} 1 \\ 0 \end{array}
 \right)  \ee
and $\tilde{F}(\theta )$ is the two-by-two matrix
\be
\tilde{F}(\theta) \eq -N(1-k' \cos \theta ) S + N k' \sin \theta \, C \period 
\ee


\subsection*{The two-by-two exponential}



 We can calculate the exponential in (\ref{defoD}) of the two-by-two
matrix $- \alpha \tilde{F}(\theta) $ in the obvious way, by diagonalizing
it, exponentiating, and then returning to the original basis. If we
define
\be \label{deflambda}
\lambda \eq  \lambda(\theta) \eq (1 - 2 k' 
\cos \theta + k'^2 )^{1/2} \comma  \ee
\bd u_{\q} (\alpha, \theta ) \eq \cosh (N \alpha \lambda) \, + 
\, \frac{1-k' \cos \theta }{\lambda}
\sinh ( N \alpha \lambda) \ed
\be \label{defvqA}
v_{\q} (\alpha, \theta ) \eq -\frac{k' \sin \theta}{\lambda} \; 
\sinh ( N \alpha \lambda ) \ee
\bd w_{\q} (\alpha, \theta ) \eq \cosh (N \alpha \lambda) \, - 
\, \frac{1-k' \cos \theta }{\lambda}
\sinh ( N \alpha \lambda) \comma \ed
then 
\be \exp[-\alpha \tilde{F}(\theta) ] \eq 
\left( \begin{array}
{cc} u_{\q} (\alpha, \theta)  & v_{\q} (\alpha, \theta) \\
v_{\q} (\alpha, \theta)  & w_{\q} (\alpha, \theta) \end{array} \right) 
\period \ee
Hence
\be \overline{D} (\cos \theta) \eq u_{\q} (\alpha, \theta ) \ee
and (\ref{resvq}) becomes
\be \label{vvq}
v_{\q} ^{\dagger}  \exp(-\alpha {\cal H} ) \, v_{\q}  \eq e^{-\mu_{\q}  \alpha} 
\, u_{\q} (\alpha,\theta_1 ) \cdots u_{\q} (\alpha,\theta_m ) \period \ee



\section{Reduced representation of $\cal H$}



\setcounter{equation}{0}

We consider some basis of the $2^m$-dimensional vector space 
$V_{\q} $ and label the vectors by $s = \{s_1, \ldots ,s_m \}$, 
where each $s_i$ takes the 
values $1$ or $-1$. We  can think of the $s_i$ as ``Ising spins''.
Thus there are $2^m$ vectors $v_{s} = v_{s}^{\q}  = $ 
$v(s_1,\ldots , v_m) $,
each of dimension $N^L$.

In  \cite{Baxter1989} we showed that we can choose the vectors $v_s$ 
so   that  $v_{\q} $ above is
\be v_{\q}  = v(1,1,\ldots , 1) \comma \ee
and
\ba {\cal H} \, v(s_1,\ldots ,s_m) & = &  \left[ \mu - N
\sum_{j=1}^m (1-k' \cos \theta_j) \,  s_j \right] 
 v(s_1,\ldots ,s_m)  + \nonumber \\ 
&& N k' \sum_{j=1}^m \sin \theta_j \, 
v(s_1,\ldots ,-s_j, \ldots, s_m) \period \ea

\noindent Defining $2^m$ by $2^m$ matrices $S_j, C_j$ by
\be
(S_j)_{s,s'} \eq s_j \prod_{n=1}^m \delta(s_n,s'_n) \comma \ee
\be
(C_j)_{s,s'} \eq  \delta(s_j,-s'_j) 
{\prod_{n=1}^m}^{\! \raisebox{-10pt}{*}}  \delta(s_n,s'_n) \comma \ee
where again the $*$ means that the term $n=j$ is excluded from the
product, we see that with respect to this basis the hamiltonian
${\cal H}$ is now
\be \label{resH}
{ H}  \eq \mu_{\q} - N \sum_{j=1}^m [(1-k' \cos \theta_j ) S_j - 
k' \sin \theta_j \, C_j ] \comma \ee
which is equation (2.20) of \cite{Baxter1989}.
This is consistent with our result (\ref{resvq}) above.

{From} (\ref{defmuq}),(\ref{resH}), $H$ is linear in $k'$.
Set 
\be \label{math1}
H = H_0 +k'  H_1 \comma \ee
$H_0, H_1$ being independent of $k'$, and
define
\be \label{Ks}
  {\kappa} (s) \eq \sum_{j=1}^m (1-s_j)/2 \comma \ee
 then $\kappa (s)$ takes the integer values
 $0, 1, \ldots ,m$. If we order the rows and columns of $H$ 
 with increasing values of $\kappa (s) $, then $H_0$ is diagonal
 and $H_1$ is block tri-diagonal, with non-zero entries only when
 $|\kappa (s) - \ \kappa (s') | \leq 1$.

{From} (\ref{resH}),  ${ H}$ is a direct sum of $m$
two-by-two matrices. Similarly, if we define the two-by-two matrix
\be U_j \eq 
\left( \begin{array}
{cc} u_{\q} (\alpha, \theta_j)  & v_{\q} (\alpha, \theta_j) \\
v_{\q} (\alpha, \theta_j)  & w_{\q} (\alpha, \theta_j) \end{array} \right) 
\comma \ee
then $\exp(-\alpha { H})$
is the direct product
\be 
\exp(-\alpha { H}) \eq e^{- r \alpha } \,
U_1 \otimes U_2 \otimes \cdots \otimes U_m \period \ee

Let $|0 \rangle$ be the $2^m$-dimensional vector whose 
elements $s$ are zero except for the element 
$s_1 = 1, s_2 = 1,  \ldots , s_m=1$, which is unity, i.e.
\be | 0  \rangle = 
\left( \begin{array} {c} 1 \\ 0\end{array} \right) \otimes 
\left( \begin{array} {c} 1 \\ 0\end{array} \right) \otimes 
\cdots  \otimes
\left( \begin{array} {c} 1 \\ 0\end{array} \right)  
\period \ee
This is the representative of the $N^L$-dimensional vector
$v_{\q} $. If $\langle 0 |$ is the transpose of  $| 0 \rangle$,
then
\be \label{vvqa}
v_{\q} ^{\dagger}  \e^{-\alpha {\cal H} } \, v_{\q}  \eq 
\langle 0  |  \e^{ - \alpha H} | 0 \rangle
\ee
and equation (\ref{vvq}) follows immediately.

The derivation of \cite{Baxter1988,Baxter1989} does not exclude the
possibility that the basis vectors $v_s$ depend on the parameter
$k'$. However, all studies for small $N, L$ agree with
the hypothesis that they are (or at least can be chosen to be) 
{\em independent}
of $k'$. This is consistent with the fact that both $\cal H$
and $H$ are linear in $k'$.


\section{The spontaneous magnetization.}




\setcounter{equation}{0}

Consider the lattice of Figure \ref{sqlatt45} and take 
$a = 0$, so all upper and lower boundary spins are fixed to 
be zero. Let $\zeta$ be the spin on a site deep inside the 
lattice. Then in the usual way we can define
the order parameters of the chiral Potts model as
\be \label{defMr}
{\cal M}_r \eq \langle \omega^{ r \, \zeta} \rangle  \ee
for $r =  1, \ldots, N-1$. Here $\langle f(\zeta) \rangle$
denotes the usual statistical mechanical average
  \be \langle f(\zeta) \rangle \eq Z_0^{-1} \, 
 \sum_{\sigma} f(\zeta)  \prod_{\langle i,j \rangle} 
  {\Ws}(\sigma_i - \sigma_j )  \prod_{\langle i,j \rangle} 
  \W(\sigma_i - \sigma_j )  \ee
for any function $f$. We take the limit when the lattice is 
infinitely large, so
$L, M \rightarrow \infty$, and $\zeta$ is infinitely far 
from the boundaries.

The ${\Ws}, \W$ products are unchanged by incrementing all spins
by one, so if we imposed toroidal boundary conditions, then it 
would be true that
\be \langle f(\zeta+1)  \rangle =  \langle f(\zeta)  \rangle   \ee
and this would imply that
${\cal M}_r = \omega^r {\cal M}_r$. Hence for $r \neq 0$ (mod $N$)
we would necessarily have ${\cal M}_r = 0$.

At high temperatures ($k' \geq 1$), this is true also for our
fixed-spin boundary conditions when we take the large-lattice
limit. However,  at lower temperaturers  ($0 < k' < 1$)
the system has ferromagnetic long-range order and ``remembers''
the boundary conditions even in the limit of $\zeta$ deep 
inside a large lattice, and 
\be 0 < {\cal M}_r < 1 \period \ee


In fact we know ${\cal M}_r$. In 1989 Albertini 
{\it et al }\cite{Albertini1989} conjectured  that
\be \label{Albconj}
{\cal M}_r \eq {(1-k'^2)}^{r (N-r)/2 N^2 } \ee
and the author was able to derive this formula in 
2005\cite{Baxter2005a,Baxter2005b}.
The method used was analytic, depending on the star-triangle
relation, functional relations and analyticity properties. 

When $N=2$ the chiral Potts model (both superintegrable and general)
reduces to the Ising model, whose partition function was obtained
by Onsager in 1944.\cite{Onsager1944} Onsager announced at a 
conference in Florence in 1949 that he and Kaufman had solved 
the spontaneous magnetization and obtained 
${\cal M}_1 = {(1-k'^2)}^{1/8}$,\cite{Onsager1949} but the first 
published derivation of that result was given by 
Yang in 1952.\cite{Yang1952}

Onsager and Yang's methods were much more algebraic, determining
the eigenvalues of the transfer matrix $T$, and certain elements
of the eigenvectors. It would be
interesting to obtain a derivation of ${\cal M}_r$ that parallels
Yang's. The object of this paper is to suggest how one may make
 progress in that  direction.

We introduce the  $N^L$ by $N^L$ diagonal matrix
 ${\cal S}_r$  with  elements
\be
({\cal S}_r)_{\sigma,\sigma'} \eq \omega^{ r \, \sigma_1} 
\prod_{j=1}^L  \delta (\sigma_j, \sigma'_j) \period \ee
Note that, for all integers ${\q} $ and $ r$,
\be \label{vS}
 {\cal S}_r v_{{\q} +r} = v_{\q}  \sep {v_{\q} }^{\dagger} {\cal S}_r = 
 v_{{\q} +r}^{\dagger} \period \ee

Because of the cylindrical boundary conditions, we can take
the spin $\zeta$ to be in any column, so we choose it to be in 
column 1. Then (\ref{defMr}) can be written
\be \label{MrWZ}
{\cal M}_r \eq W/ Z_0 \comma \ee
where
\be W \eq   u_0^{\dagger} \, T^j {\cal S}_r
 T^{M-j} u_0 \sep Z_0 =  u_0^{\dagger} \, T^M u_0 \comma \ee
$j$ being the number of rows below $\zeta$.


{From} (\ref{vu}) and (\ref{ZtZ}),
\be \label{Weq}
 W = N^{-1} \sum_{\q ,\p  =0}^{N-1} v_{\q} ^{\dagger}  T^j {\cal S}_r
 T^{M-j} v_{\p} \sep Z_0 = N^{-1} \sum_{{\q} =0}^{N-1} 
v_{\q} ^{\dagger} T^M v_{\q}  \period \ee
Since $R$ commutes with $T$ and
\be \label{RS}
R {\cal S}_r = \omega^{-r} {\cal S}_r R \comma \ee
it follows from (\ref{RVq}) that the first summand in (\ref{Weq})
vanishes unless $\p = \q +r$, so
\be \label{Weq2}
 W = N^{-1} \sum_{\q =0}^{N-1} v_{\q} ^{\dagger}  T^j {\cal S}_r
 T^{M-j} v_{{\q} +r} \comma \ee
 interpreting ${\q} +r$ as ${\q} +r$ to modulo $N$.

 For $0 < k' < 1$ and $L$ is large, the $N$  largest 
eigenvalues of  $T$ are asymptotically degenerate, their  
ratios being of the form $ 1 + O(\e^{-L \nu})$, $\nu$ being a measure
of the interfacial tension. However, there is one and only one
of these eigenvalues in each of the vector spaces $V_{\q} $, for 
${\q} = 0,\ldots, N-1$.

Since $T$ and $\cal H$ commute and $\cal H$ is hermitian, the 
eigenvectors $\psi_{\q} $ corresponding to these eigenvalues are 
unitary, so
\be \psi_{\q}^{\dagger} \psi_{{\p} } \eq \delta_{p,q} \period \ee


\subsection*{Asymptotic degeneracy}


In each sub-space $V_{\q} $ there is single largest eigenvalue 
$\Lambda_{\q} $ of the transfer matrix $T$ and these eigenvalues are
{\em asymptotically degenerate}, in the sense that for large $L$
there is a common value $\Lambda$ such that
\be \label{asymp}
 \Lambda^{-1} \Lambda_{\q}  \eq   1 + O(\e^{-L s_{\q}  }) \comma \ee
i.e. the ratios of the $\Lambda_{\q} $ approach unity exponentially rapidly.

This can be seen by considering the series expansion of the
eigenvector $\psi_q$ in powers of $k'$.  Since $T$, ${\cal H}$ commute, 
we  can look at the eigenvectors of $\cal H$, corresponding
to the most negative (ground state) eigenvalue.

When $k'=0$, ${\cal H} = {\cal H}_0$, where
\be {\cal H}_0 =  -2 \sum_{j=1}^L \sum_{n=1}^{N-1} {\Z}_j^n 
{\Z}_{j+1}^{-n}/(1-\omega^{-n})
\period \ee
This is diagonal, with minimum eigenvalue $-2L$, when
all the $L$ spins are equal. Thus from (\ref{defua}), 
$u_0, \ldots, u_{N-1}$ are ground state eigenvectors.

We can start from one of these eigenvectors and use
standard  linear perturbation theory to develop a
series expansion for the eigenvector of ${\cal H}$, starting from
the initial eigenvector $u_a$. This 
entails changing successively more of the spins from value $a$ to 
some other value. It will work until all of the spins are changed, 
when for the first time we come to another of the eigenvectors
of ${\cal H}_0$. At that stage, and only at that stage,
 one would have to resolve the 
degeneracy of the initial eigenvalues.
This means that naive perturbation theory works to order ${k'}^L$.
The calculation only depends on $a$ in so far as it
involves the differences (mod $N$) of the $L$ spins from $a$.
Thus to this order the eigenvalue is independent of the
initial choice of $a$. This is true also of the eigenvalues of $T$, 
so $\Lambda_{\q}  = \Lambda$, $\Lambda$ being the common 
eigenvalue, in agreement with (\ref{asymp}).

Also, if $\psi'_a$ is this near-eigenvector, 
then
\be { \psi'_a}^{\dagger}  u_b =   \xi \delta_{a,b}  \comma \ee 
where $\xi $ is independent of $a$ and $b$, and to this order
the actual eigenvectors are
\be \psi_{\q}  = N^{-1/2} \, \sum_{a=0}^{N-1} 
\omega^{-a {\q} } \psi'_a \period \ee
it follows that, for all ${\q} $,
\be \psi_{\q} ^{\dagger}  v_{\q}  = \xi \period \ee

In the limit of $j, M-j, L$ large we can replace $T^j$ in (\ref{Weq}), 
(\ref{Weq2}) by   $\psi_{{\q} } \Lambda^j \psi_{\q} ^{\dagger}$ (with the 
appropriate value of ${\q} $),  and  $T^{M-j}$ by 
 $\psi_{{\q} } \Lambda^{M-j} \psi_{\q} ^{\dagger}$, giving
  \be \label{asymp2}
  v_{\q} ^{\dagger}  T^j {\cal S}_r  T^{M-j} v_{{\q} +r}
= \e^{M \Lambda} \xi^* \xi \,
 \psi_{\q} ^{\dagger} {\cal S}_r \psi_{{\q} +r} \sep 
 v_{\q} ^{\dagger} T^M v_{\q}  =  \e^{M \Lambda} \xi^*\xi  \ee
 $\xi^*$ being the complex conjugate of $\xi$ and 
 \be \label{indepq}
 \psi_{\q} ^{\dagger} {\cal S}_r \psi_{{\q} +r} = \; \; {\rm independent \; of} 
 \; \; {\q}  \period \ee
 Thus $W, Z_0$ are the two expressions in (\ref{asymp2}),
 respectively, and
 \be \label{calcM}
  {\cal M}_r \eq  \psi_{\q} ^{\dagger} {\cal S}_r \psi_{{\q} +r} 
 \period \ee



\subsection*{Expressions in terms of $\cal H$}


Rather than continue to work with the transfer matrix $T$, we find it
convenient to instead use the negative exponential of
the hamiltonian
and to replace $T^j, T^{M-j}$ in 
(\ref{Weq2}) by $\e^{-\alpha {\cal H}}$,  $\e^{-\beta {\cal H}}$,
and $T^M$ in (\ref{Weq}) by  $\e^{-\alpha {\cal H}}$ (with a different 
$\alpha$), making them 
\be W \eq N^{-1} \sum_{{\q} =0}^{N-1} \tilde{W}_{\q, \p}  \sep Z_0 \eq N^{-1} 
\sum_{q=0}^{N-1}  \tilde{Z}_q \comma \ee
where now, setting $\p = {\q} +r$,
\be \label{Weq4}
 {\widetilde{W}}_{{\q} , \p }  \eq  {\widetilde{W}}_{{\q} , \p}(\alpha, \beta , x) \eq 
 v_{\q} ^{\dagger} \e^{-\alpha {\cal H}} \e^{-\rho {\cal J}}
  {\cal S}_r
\e^{-\beta {\cal H}}  v_{\p } \ee
\be \label{deftZ}
 \tilde{Z}_{\q}  \eq   \tilde{Z}_{\q}  (\alpha) \eq 
v_{\q} ^{\dagger} \e^{-\alpha  {\cal H}} v_{\q}    \ee
and 
\be \label{defx}
 x = \e^{- 2 N  \rho} \period \ee

We have   introduced the matrix factor  $ \e^{-\rho {\cal J}}$
immediately pre-multiplying   ${\cal S}_r$ in (\ref{Weq4}).
Here 
\be  {\cal J}  \eq {\cal H}_0  +L(N-1) I   \ee
is a diagonal matrix  whose entries are 
 $0, 2 N, 4 N, \ldots ,$ $2 N [(N-1)L/N]$. Hence 
 $\widetilde{W}_{{\q} , \p }(\alpha, \beta,x)$ is a  polynomial
 in $x$ of degree  $[(N-1)L/N]$. This naturally manifests itself in the
 following working and provides a useful check against errors.

 We can think of these $\tilde{Z}_{\q} $, $\tilde{W}_{{\q} , \p }$ as 
 hamiltonian partition functions. They are rather simpler than the original
 partition functions to work with.

 When $\rho \rightarrow + \, \infty$, then $x \rightarrow 0$
 and    $ \e^{-\rho {\cal J}} \rightarrow v_{\q}  {v_{{\q} }}^{\dagger}$, 
 so, using (\ref{vS}),
  \setlength{\jot}{0.2cm}
 \ba  \label{Winf}
 {\widetilde{W}}_{{\q} , \p } (\alpha, \beta, 0)  & = & 
  v_{\q} ^{\dagger} \e^{-\alpha {\cal H}} v_{\q}  \, {v_{\p }}^{\dagger}
   \e^{-\beta {\cal H}}  v_{\p} \nonumber \\
& = &     \tilde{Z}_{\q}  (\alpha)  \, \tilde{Z}_{\p }(\beta) 
   \comma \ea
\be \label{W0}
 {\widetilde{W}}_{{\q} , \p }(\alpha, 0, x) \eq   \tilde{Z}_{\q} (\alpha)  \
\sep  {\widetilde{W}}_{{\q} , \p }(0,\beta, x ) \eq   \tilde{Z}_{\p }(\beta)  
 \period  \ee
These relations also provide useful checks on our subsequent
calculations.

Because ${\cal H}, T$ commute, they have the same ground-state 
eigenvectors $\psi_{\q} $. In the limit when $\rho=0, $ and $\alpha, 
\beta, L  \rightarrow \infty$, we obtain
\bd 
\widetilde{W}_{{\q} , \p }(\alpha, \beta, 1) = \e^{-(\alpha + \beta ) \Lambda}
\xi^* \xi \, \psi_{\q} ^{\dagger} {\cal S}_r \psi_{\p} \comma \ed
\be \tilde{Z}_{\q} (\alpha ) =  \e^{-\alpha \Lambda } 
\xi^* \xi \period \ee
So from (\ref{indepq}), (\ref{calcM}),
\be \label{exprM}
 {\cal M}_r \eq  \lim_{\alpha, 
\beta, L  \rightarrow \infty} 
\frac{ \widetilde{W}_{ {\q}  ,\p }(\alpha, \beta, 1) 
}{(\tilde{Z}_{\q} ( 2\alpha )  \tilde{Z}_{\p}( 2 \beta ) )^{1/2}}
\ee
for any 
${\q} , \p $ such that $0 \leq \q, \p <N$ and $\p  = {\q} +r$,  mod $N$.

{From} (\ref{vvq}) and (\ref{deftZ}),
 \be \label{tilZ}
  \tilde{Z}_{\q} (\alpha )  \eq e^{-\mu_{\q}  \alpha} 
\, u_{\q} (\alpha,\theta_1 ) \cdots u_{\q} (\alpha,\theta_m ) \period \ee

It remains to calculate $\widetilde{W}_{{\q} , \p }(\alpha, \beta, x) $.
We have not done this, but the rest of this paper
is concerned with presenting a conjecture for it 
as a determinant of dimension not greater than
$(N-1)L/N$. This expression agrees with the known 
$N=2$ result for the Ising model, and indeed is a fairly 
immediate generalization of that result.
It has the properties (\ref{Winf}),
(\ref{W0}), and has been extensively tested numerically for small 
values of $N, L$.


\subsection*{Expressions in terms of $H$}



First we remark that if $v \in V_{{\q} +r}$ and $v' = {\cal S}_r v$, then from
(\ref{RVq}), $R v' = \omega^{{\q} } \, v' $, so $v'$ is a 
candidate for the sub-space $V_{{\q} }$. However, in general it 
does not lie within this sub-space. Even so, we can define a matrix
${\cal S}_{\rm red}^r$ of dimension $m_{\q} $ by $m_{{\q} +r}$ by
\be \label{defSred}
\left( {\cal S}_{\rm red}^r\right) _{s,s'} \eq \left( v_s^{\q}  
\right)^{\dagger} S_r v_{s'}^{{\q} +r} \period \ee
These elements depend on $N, L, {\q} , r$. They are of course 
independent of $\alpha$ and $\beta$.  From our remarks at the 
end of section 4 that we expect the $v_s$ to be independent of 
$k'$, the same must be true of  the  elements  of 
${\cal S}_{\rm red}^r$.

We can then write (\ref{Weq4}) as
\setlength{\jot}{3mm}
\ba \label{Weq5}
{\tilde{W}}_{{\q} , \p } & =
 &   \langle 0  |  \e^{-\alpha H}  \e^{-\rho J} 
 {\cal S}_{\rm red}^r  \e^{-\beta H'} 
 | 0 \rangle 
 \comma  \\
\tilde{Z}_{\q}   & = &    \langle 0  | 
  \e^{-\alpha H } | 0  \rangle 
\comma \nonumber \ea where
$H'$ is the $H$ of  (\ref{resH}),(\ref{math1}) but with 
$\q$ replaced by $\q+r$, and 
 \be  J  = H_0
 + L(N-1) I  =  N \sum_{j=1}^m (I-S_j)  \ee
 is the diagonal matrix with elements $2 N \kappa (s)$ in position
 $(s, s)$.
 
Let
\be \tilde{u}_{\q} (\, 1,\alpha,\theta ) = u_{\q} (\alpha,\theta )\sep 
\tilde{u}_{\q} (-1,\alpha,\theta ) = v_{\q} (\alpha,\theta )\period \ee
and set
\be \p = {\q} +r \sep  m' = m_{\p} \sep \mu' = \mu_{\p}  \sep 
\theta'_j = \theta(\p, j) \period \ee   
Then we can write these equations  more explicitly as
\setlength{\jot}{1mm}
\ba  \label{Weq6}
\widetilde{W}_{{\q} , \p }(\alpha, \beta, x)  & = &   e^{-\alpha \mu -\beta \mu'} 
\, \sum_{s,s'}  \tilde{u}_{\q} (s_1,\alpha,\theta_1 ) \cdots 
{\tilde{u}}_{\q} (s_m,\alpha,\theta_m ) \; 
{\mbox{\large $ \times$ } }   \nonumber \\
&& x^ {{\kappa} (s) }  \left( {\cal S}_{\rm red}^r\right) _{s,s'} 
\tilde{u}_{\p}(s'_1,\beta,\theta'_1 ) \cdots 
{\tilde{u}}_{\p}(s'_m,\beta,\theta'_{m'} ) \comma \ea
\noindent and
\be \label{resZt}
\tilde{Z}_{\q}(\alpha)    =  e^{-\alpha \mu } 
\, u_{\q} (\alpha,\theta_1 ) \cdots u_{\q} (\alpha ,\theta_m ) \period  \ee
  The non-zero
 elements $(s,s')$ of the $2^m$ by $2^{m'}$ 
 matrix $ {\cal S}_{\rm red}^r$ 
  satisfy $ {\kappa} (s) =  {\kappa} (s')$.
 If we also order the rows and columns of   
 $ {\cal S}_{\rm red}^r$ in increasing value of $\kappa (s)$, then
 this matrix is block-diagonal.
 
 We do not have a direct derivation of  $ {\cal S}_{\rm red}^r$, though of 
 course it can be calculated numerically for small values of $N, L$ from
 (\ref{defSred}). In principle it can be calculated from our conjecture
 (\ref{conj}) below. If $\bf s$ is the $m$ by $m'$ 
 diagonal blocks of $ {\cal S}_{\rm red}^r$ in the block
 $\kappa (s) = \kappa (s') = 1$,  $\bf h$ is the corresponding
 $m$ by $m$ block of $H_1$, and  ${\bf h}'$  the $m'$ by $m'$
 block of $H'_1$,
 then this conjecture
 implies that the double commutator 
 $\bf h{\cdot}h{\cdot}s-2 \,  h{\cdot}s{\cdot}h'+
 s{\cdot}h'{\cdot}h'$ is of rank one. This was a key 
 initial encouraging observation in our search for the expression  (\ref{conj}).


\section{The orthogonal matrix $B$ }




\setcounter{equation}{0}

Before stating our conjecture, we define an $m$ by $m'$ 
real orthogonal matrix $B = B_{\q \p }$ whose elements involve 
the $\theta_1, \ldots,
\theta_m$ defined by  (\ref{defP}), (\ref{deftheta}), as well
as the $\theta'_1, \ldots,\theta'_{m'}$ defined similarly, but
with ${\q} $ replaced  by $\p$ and $m$ by $m'$. We must have
${\q}  \neq \p$.

We define $B = B_{\q ,\p }$ to be the matrix with elements
\be \label{defB}
B_{i,j} \eq f({\q} , \p ,i) f(\p ,{\q} ,j)/(\cos \theta_i - \cos \theta'_j)
\comma \ee
where we choose  the functions $f({\q} , \p ,i), f(\p ,{\q} ,j)$ to ensure
that
\be  B^T B = I \;  \; {\rm if \; \;} m  \geq m' \sep
  B B^T = I \;  \; {\rm if \; \;} m  \leq m' \comma \ee
$I$ again being the identity matrix, of dimension $\min (m,m')$.


\subsection*{The case $\q < \p$}


{From} (\ref{defm}) , if ${\q}  < \p $, then $ m \geq m'$ and we want
$B^T B = I$. {From}  (\ref{defB}),
\bd (B^T B)_{i,j} \eq \sum_{n=1}^{m}
 \frac{f(\p ,\q ,i) f({\q} , \p ,n)^2 f(\p ,{\q} ,j)}{((\cos \theta_n - 
\cos \theta'_i) (\cos \theta_n - \cos \theta'_j))} \ed
\be \label{BBT}
\eq \frac{f(\p ,{\q} ,i) f(\p ,{\q} ,j)}{\cos \theta'_j - \cos \theta'_i}\; 
\sum_{n=1}^m \left\{ \frac{ f({\q} , \p ,n)^2}{\cos \theta'_i - \cos \theta_n} -
\frac{ f({\q} , \p ,n)^2}{\cos \theta'_j - \cos \theta_n} \right\}  \ee
for $i \neq j$.

We want the RHS of (\ref{BBT}) to vanish for $i \neq j$.
Consider the functions
\be \tilde{P}_{\q} (c) =  \prod_{i=1}^m (c-\cos \theta_i)
= N^{-L} (c+1)^m \, P {\textstyle \left( \frac{c-1}{c+1} \right) } \comma \ee
\be \label{defFc}
{\cal F}(c) \eq 
 \sum_{n=1}^m  \frac{ f({\q} , \p ,n)^2}{c- \cos \theta_n } \period \ee
The first is a known function, given by (\ref{defP}) and
(\ref{deftheta}), the second is of the form
${\cal R}_{\q} (c)/\tilde{P}_ {\q} (c)$, ${\cal R}_{\q} (c)$ being a polynomial
of degree $m-1$. We want there to exist constants 
$\gamma, \gamma'$ (dependent on ${\q} ,  \p $) such that
\be \label{formFc}
{\cal F}(c) \eq \gamma' + \gamma \tilde{P}_{\p}(c)/\tilde{P}_{{\q} }(c)
\comma \ee
since then ${\cal F} (\cos \theta'_i ) = {\cal F}(\cos \theta'_j ) = \gamma'$
and the RHS of (\ref{BBT}) vanishes.
This implies that
\be \label{req}
{\cal R}_{\q} (c) = \gamma' \tilde{P}_{\q} (c) +\gamma 
 \tilde{P}_{\p}(c) \period \ee

{From}  (\ref{defm}), $m$ and $m' = m_{\p}$ differ by at most
one, so $m'+1 \geq m \geq m'$.
Whether $m= m'$ or $m= m'+1$, we can always choose
$\gamma'$ to ensure that the RHS of (\ref{req}) is a polynomial 
of degree $m-1$. Then the equation defines ${\cal R}_{\q} (c)$
(to within the factor $\gamma$) and the 
parameters $f({\q} , \p ,n)$. 

{From}  (\ref{defFc}), $f({\q} , \p ,n)^2$ is the  residue of ${\cal F}(c)$ at 
the pole   $c = \cos \theta_n$, so from (\ref{formFc})
\be \label{calcfqn}
f({\q} , \p ,n)^2 \eq \gamma \, \tilde{P}_{\p}(\cos \theta_n)/
\Delta_{\q}  (\cos \theta_n) \comma \ee
where
\be \Delta_{ {\q} } (c) \eq \frac{d}{dc} \tilde{P}_{{\q} }(c)\period \ee


For given ${\q} , \p $, this determines $f({\q} , \p ,i)$ to within a 
factor independent  of $i$ (but possibly dependent on ${\q} $ and $\p$). 
To determine this  factor we need to consider the case 
when $i=j$ in the first of the equations  (\ref{BBT}), which gives
\be f({\q} , \p ,i)^2  \, G(\cos \theta'_i) = 1 \comma \ee
where
\be G(c) = \sum_{n=1}^m f({\q} , \p ,n)^2/(c - \cos \theta_n)^2 
\period \ee
{From} the equations above,
\be G(c) = - \frac{d}{dc} F(c) 
= - \gamma \frac{d}{dc} \, \frac { \tilde{P}_{\p} (c)}
{ \tilde{P}_{{\q} } (c)}
 \period \ee
Since $ \tilde{P}_{\p} (\cos \theta'_i) = 0$, this gives
\be  \label{calcfqpi}
f(\p ,{\q} ,i)^2 = \frac{1}{G(\cos \theta'_i)} =  - \frac{\tilde{P}_{\q}
 (\cos \theta'_i) }{\gamma \,\Delta_{\p}(\cos \theta'_i )}
\period \ee

The parameter $\gamma$ is at our disposal. We observe numerically 
that for small values of $n$ and $L$ we
can ensure that $f({\q} , \p ,n)^2$, $f(\p ,{\q} ,i)^2$ are real and positive by
choosing 
\be \label{setgamma}
\gamma = 1 \period \ee
We can then take $f({\q} , \p ,n)$, $f(\p ,{\q} ,i)$ to be positive, for all $n, i$.
The matrix $B$ is then defined by (\ref{defB}), (\ref{calcfqn}),
(\ref{calcfqpi}), (\ref{setgamma}). It is real and has the 
orthogonality property $B^T B = I$. 
If $m= m'$ this implies $B B^T = I$.


\subsection*{The case $\q > \p$}


We can combine (\ref{calcfqn}), (\ref{calcfqpi}) into a single
formula by defining
\be
\epsilon({\q} , \p ) = 1 \; \; {\rm if} \; {\q}  <  \p \sep 
\epsilon({\q} , \p ) = \; -1 \; \; {\rm if} \; {\q}  > \p \period \ee
Then both equations are contained in
\be \label{Bsumm}
f({\q} , \p ,i) \eq \left[ \epsilon({\q} , \p )  \, \tilde{P}_{\p}(\cos \theta_i)/
\Delta_{\q}  (\cos \theta_i) \right]^{1/2} \comma \ee
for ${\q}  \neq \p$.

We can now extend the formula (\ref{defB}) to all ${\q}  \neq \p$. It is 
readily observed that
\be B_{\p ,{\q} } = - B_{{\q} , \p }^T \period \ee
We have just established that $B^T B = I$ if ${\q}  <  \p$. It follows that
$B B^T = I$ if $ {\q}  >  \p$ (which implies $m \leq m'$). This is the desired 
orthogonality property.

We remark that we have only conjectured (based on numerical
calculations) that the RHS of 
(\ref{Bsumm}) is real and can be chosen positive. If this
were to fail the above formulae would still apply, but $B_{{\q} \p }$
would be a complex orthogonal matrix.



\subsection*{The matrix  $E$}


 We shall also need the $m$ by $m$  diagonal matrix $E_{\q \p}$,
 with entries 
 \be
  [E_{\q ,\p}]_{i,j}  \eq e({\q} ,\p,i) \; \delta_{i,j}  \comma  \ee
where 
  the function $e({\q} , \p ,i)$ is defined as follows, for 
 $ 0 \leq {\q} , \p  < N$:
 \ba  \label{defgfns}
 e(\q, \p, i) = &   \sin \theta_i   \;  & {\rm if} \; \; {\q}  < \p  \; {\rm and } \;
 m > m'  \nonumber \\
=  &  \tan ( \theta_i /2 ) \;  &  {\rm if} \; \; {\q}  < \p   \; {\rm and } \;
 m  =  m'  \nonumber \\
=  &  1/\sin \theta_i   \; & {\rm if} \; \; {\q}  > \p   \; {\rm and } \;
 m <  m'   \\
 =  &  \cot ( \theta_i /2 ) \;   & {\rm if} \; \; {\q}   > \p   \; {\rm and } \;
 m  =  m' \period  \nonumber \ea
Since $ m-1 \leq m' \leq m $ if ${\q}  < \p $, and 
 $ m + 1 \geq m' \geq m $ if ${\q}   >  \p $, these equations cover all cases;
  $\theta_i = \theta({\q} ,i)$ is again as defined in (\ref{deftheta}). The
function $e(\p ,{\q} ,i)$ is defined similarly, but with ${\q} , \p $ interchanged
and $\theta_i$ replaced by $\theta'_i= \theta(\p ,i)$.


\section{The conjecture for $W$}


\setcounter{equation}{0}

We return to considering the $\widetilde{W}_{{\q} ,\p }$ of equations
(\ref{Weq4}), (\ref{Weq5}) and  (\ref{Weq6}).
Based on the calculation for the Ising model,\cite[eq.7.9]{Baxter2008}
we conjecture that
\be \label{cnjW}
\widetilde{W}_{{\q} ,\p } (\alpha, \beta, x) \eq   \tilde{Z}_{\q}  (\alpha)  \,
 \tilde{Z}_{\p}(\beta) \,  {\cal D}_{{\q} ,\p } (\alpha,\beta)
\comma \ee
where ${\cal D}_{{\q}, \p } (\alpha,\beta)$ is the $m$ by $m$ determinant
\be \label{conj}
{\cal D}_{{\q} , \p } (\alpha,\beta) \eq  \det [ I _m - x 
X_{\q} (\alpha) E _{{\q} ,\p } \, B _{{\q} ,\p } X_{\p}(\beta)  E_{\p ,{\q} } 
  \, B_{\p ,{\q} }  ] \ee
or equivalently the $m'$ by $m'$ determinant
\be \label{conj2a}
{\cal D}_{{\q}  ,\p } (\alpha,\beta) \eq  \det [ I _{m'} - x 
X_{\p} (\beta) E_{{\p} ,\q } \, B _{{\p} ,\q } X_{\q}(\alpha)  E_{\q ,{\p} } 
  \, B_{\q ,{\p} }  ]  \period \ee
Again $I_m$ is the identity matrix, of dimension $m$ and   
$X_{\q} (\alpha)$ is the diagonal $m$ by $m$ matrix whose 
entry in position $(i,j)$ is
\be 
[X_{\q} (\alpha)]_{i,j} \eq \frac{v_{\q} (\alpha, \theta_j)}
{u_{\q} (\alpha, \theta_j)} \, \delta_{i,j} \ee

Note from (\ref{defB})  that each function
$f({\q} , \p ,i),  f(\p ,{\q} ,j)$  occurs twice (i.e. as its square)  in
(\ref{conj})  and (\ref{conj2a}), so the choice of the square roots
in  (\ref{Bsumm}) is in fact irrelevant.

 {From} (\ref{defvqA}) and (\ref{Weq4}), $v_{\q} (\alpha, \theta) = 0 $ and 
 $\tilde{Z}_{\q} (\alpha) = 1$ if $\alpha = 0$, so (\ref{cnjW}) does indeed 
 have the properties (\ref{Winf}), (\ref{W0}). It is a
 fairly  immediate generalization
 of eqn. (7.7) of \cite{Baxter2008} and has been tested to high numerical 
 accuracy (60 digits or more) for arbitrary $k', \alpha, \beta$ and all 
 $N, L, p, q$ such that $2 \leq N $, $ 3 \leq L $, $ N+L \leq 10$.  We 
 {\em conjecture} that it is true for all $N, L, p, q, x, \alpha, \beta$.
 
 


 \subsection*{Consequences}


Define \be a_{\q ,j}  \eq    \{ 1- k' \e^{\i \theta_j} \}^{1/2} \sep
 b_{\q ,j}  \eq    \{ 1- k' \e^{-\i \theta_j} \}^{1/2}  \comma \ee
 where $\theta_1, \ldots, \theta_m$ are given by (\ref{deftheta}). They 
 depend on $\q$. Again the function $m_{\q}$ is defined by 
 (\ref{defm}) for $0 \leq \q < N$, and $m = m_{\q}$, $m' = m_{\p}$.
 
 Then from (\ref{deflambda}), 
 \bd
 \lambda_j =  \lambda(\theta_j ) = 
 (1-2k' \cos \theta_j +k'^2)^{1/2} = a_{\q ,j} b_{\q ,j} \comma \ed
 so  from (\ref{tilZ}) and  (\ref{defvqA}),
 \be \lim_{\alpha \rightarrow \infty}  \; \frac{\tilde{Z}_{\q} ( \alpha)^2}{
 \tilde{Z}_{\q} ( 2\alpha) } 
=  \prod_{j=1}^m \frac{(a_{\q ,j}+b_{\q ,j})^2}{4 \, a_{\q ,j} b_{\q ,j}} 
\period \ee
Also define quantities $x_{\q,j}$,
not to be confused with the $x$ of (\ref{defx}),  by
\be  x_{\q, j}  =    \lim_{\alpha \rightarrow \infty} 
\frac{v_{\q} (\alpha, \theta_j)}{u_{\q} (\alpha, \theta_j)}  =
\frac{- \, k' \sin \theta_j }{\lambda_j +1 - k' \cos \theta_j } \period \ee
Then \be 
x_{\q, j}  =  \i \; \frac{ b_{\q,j} - a_{\q, j} }{ b_{\q,j}  + a_{\q, j}} \ee
and 
\be
\lim_{\alpha \rightarrow \infty}  \; \frac{
 \tilde{Z}_{\q} ( 2\alpha) }{\tilde{Z}_{\q} ( \alpha)^2} = 
 \prod_{j=1}^m \left( 1+x_{\q,j}^2 \right)\period \ee
 
 Let
 \be X_{\q} = \lim_{\alpha \rightarrow \infty}  \; X_{\q}(\alpha) \comma \ee
 so it is the diagonal matrix with diagonal elements $x_{\q,j}$. Taking the 
 limits $\alpha,  \beta \rightarrow + \infty$ and setting $x=1$, it follows 
 from (\ref{exprM}), (\ref{cnjW}) that   if $\p = \q+r$  to modulo  $N$, then
  \renewcommand{\arraystretch}{0.2}
  
 \be \label{resultM}
 {\cal M}_r \eq \lim_{L \rightarrow \infty}  \; \begin{array} {c}  \det ( I _{m} - 
X_{\q}  E _{{\q} ,\p } \, B _{{\q} ,\p } X_{\p} E_{\p ,{\q} } 
  \, B_{\p ,{\q} }  ) \\[5pt]
  \cline{1-1} \\
  \{ \det (I_m + X_{\q}^{2} )\, \det (I_{m'}+ X_{\p}^2 ) \}^{1/2} 
 \end{array} 
\ee
where  $0 \leq \q, \p <N$ and $ 0 < r < N$.

 \renewcommand{\arraystretch}{1.0}
 
 We have not been able to evaluate the RHS of (\ref{resultM}) analytically. 
 Even for the $N \! = \! 2 \, $  Ising case discussed in \cite{Baxter2008}, we 
 do  not directly evaluate (\ref{resultM}), but rather the expression in terms 
 of square roots of $L$ by $L$ determinants that leads in that case to 
 (\ref{resultM}).\footnote{We do this by writing ${\cal M}_r^2$ as the 
 determinant of a Toeplitz matrix and using Szeg{\H o}'s theorem.}
 
 We have conducted numerical experiments  for various values 
 of $N, \q, \p$   and $k'$, 
 and observed that as $L \rightarrow \infty$ the expression on the 
 RHS of (\ref{resultM})
 does indeed approach the known result (\ref{Albconj}), the error for finite 
 $L$ being of the order of $k'^L$ or smaller.


\section{Summary}

\setcounter{equation}{0}

We have defined the hamiltonian partition functions 
$\widetilde{W}_{\q,\p} (\alpha, \beta,x)$, $\tilde{Z}_{\q}(\alpha)$ by
(\ref{Weq4}), (\ref{deftZ}) and shown that the spontaneous magnetization 
${\cal M}_r$ of the superintegrable chiral Potts model is given by
 (\ref{exprM}). 
For the general solvable chiral Potts model, ${\cal M}_r$ is independent of 
the rapidities\cite[p.7]{Baxter2005b}. The superintegrable model is obtained
from the general by a special choice of the rapidities 
($k'$ being the same), \cite[p.5]{Baxter1989}
so ${\cal M}_r$ is the same for both.\footnote{Note that the $\q, \p$ of 
this paper are {\em not} rapidities.}


By taking the hamiltonian limit of the results of \cite{Baxter1989},
we show that $\tilde{Z}_{\q}(\alpha)$  is given by (\ref{resZt}). We then 
conjecture that $\widetilde{W}_{\q,\p} (\alpha, \beta,x)$ is given in terms
an $m$ by $m$ determinant by (\ref{cnjW}). This is a natural 
generalization of the known result for the special case $N = 2$, 
i.e. the Ising model.\cite{Baxter2008}

If this is true (and all the numerical evidence suggests that it is) this is
a huge simplification, reducing the problem from exponential complexity
to comparitively small polynomial complexity. Even so, we have not
been able to make the final step and to obtain ${\cal M}_r$ from
(\ref{resultM}). We already know\cite{Baxter2005a,Baxter2005b}
 that ${\cal M}_r$ is given by (\ref{Albconj}), but it would be 
 interesting to obtain it by this more algebraic route. The matrices
 $B_{pq}$ and (for $m \geq m' $) $ {\cal D}_{p,q}(\alpha, \beta) B_{p,q}$ are 
 Pick matrices.\cite{Agler2002}

So there remain two things to do: to prove the conjecture
(\ref{cnjW}) and to evaluate the limit (\ref{resultM}). The first is an 
algebraic problem, the second an analytic one. The fact that 
(\ref{cnjW}) contains the additional parameters $\alpha, \beta, x$
should be helpful in establishing it.


\section{Acknowledgement}

The author is grateful to Helen Au-Yang for helpful comments and for 
pointing out a number of typographical errors in this and the preceding 
paper \cite{Baxter2008}.



 \end{document}